\documentclass[%
superscriptaddress,
 amsmath,amssymb,
11pt,
onecolumn
]{revtex4-2}
\DeclareMathOperator{\sinc}{sinc}

\usepackage{graphicx}
\usepackage{hyperref}

\begin{document}
\title{Time-to-space ghost imaging}
\author{Dmitri B. Horoshko}\email{dmitri.horoshko@univ-lille.fr}
\affiliation{Univ. Lille, CNRS, UMR 8523 - PhLAM - Physique des Lasers Atomes et Mol\'{e}cules, F-59000 Lille, France}
\affiliation{B. I. Stepanov Institute of Physics, NASB, Nezavisimosti Ave 68, Minsk 220072 Belarus}

\begin{abstract}
Temporal ghost imaging is based on the temporal correlations of two optical beams and aims at forming a temporal image of a temporal object with a resolution, fundamentally limited by the photodetector resolution time and reaching 55 ps in a recent experiment. For further improvement of the temporal resolution, it is suggested to form a spatial ghost image of a temporal object relying on strong temporal-spatial correlations of two optical beams. Such correlations are known to exist between two entangled beams generated in type-I parametric downconversion. It is shown that a sub-picosecond-scale temporal resolution is accessible with a realistic source of entangled photons.
\end{abstract}

\maketitle

In spatial ghost imaging \cite{Belinskii94,Pittman95,Bennink02,Gatti04}, the image of a sample is formed by detecting two correlated optical beams: the test beam passing through the sample is detected by a single-pixel detector, while the reference beam is detected by a camera with a high spatial resolution. Neither of the two detection records is sufficient for building an image of the sample, which appears only in the correlation function of the two records and relies on strong spatial ($x-x$) correlations between the test and reference beams.

In temporal ghost imaging (TGI) \cite{Shirai10,Ryczkowski16,Denis17,Wu19} the image of a temporal object, whose transmittivity changes with time, is formed in a similar manner by detecting two temporally  correlated optical beams: the test beam passing through the sample is detected by a single-temporal-pixel detector, while the reference beam is detected by a fast detector with a high temporal resolution.  Again, the image appears only in the correlation function of the two recorded data sets and relies on strong temporal ($t-t$) correlations between the test and reference beams. Both the spatial and temporal ghost imaging techniques attract much attention due to their inherent insensitivity to the distortion that may occur between the object and the single-pixel detector, allowing one to form high-resolution images in a strongly scattering medium, i.e. in optical coherence tomography \cite{Amiot19,Huyan22} or ultra-high frequency signal transmission \cite{Wang22}. In addition, ghost imaging based on photon pairs admits  dual-color imaging \cite{Aspden15} with a reference beam at a wavelength below 1 $\mu$m, detected with high quantum efficiency and resolution by silicon detectors, and a longer-wavelength test beam, carrying information on the object transmittivity in the infrared. The temporal resolution of a temporal imaging system is determined by the response time of the fast detector and its best value reported is 55 ps \cite{Ryczkowski16}. This value is already at the limit of the temporal resolution of photodetectors and its improvement is possible by temporal magnification in the reference arm \cite{Ryczkowski17} and by decreasing the correlation time of the beams by employing, e.g. a fiber laser as a source \cite{Wu20}.

The main idea of this Letter is to combine two approaches described above and form a spatial ghost image of a temporal object relying on strong temporal-spatial ($t-x$) correlations between the test and reference beams. Such a technique uses spatial measurement of the transverse intensity distribution of the reference beam and thus avoids the limitation of the detector speed. The temporal resolution can reach hundreds of femtoseconds in a realistic example, realizing a dual-color imaging scenario with the spectral ranges of the reference and test beams 765-987 nm and 1.15-1.74 $\mu$m, respectively.

\begin{figure*}[ht]
\centering
\fbox{\includegraphics[width=\textwidth]{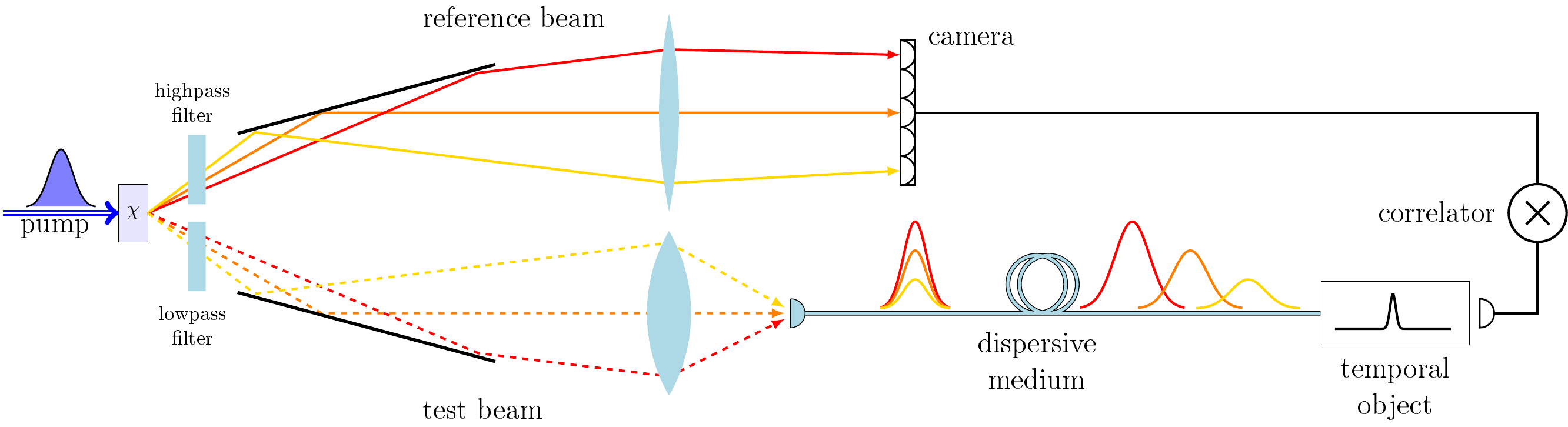}}
\caption{Scheme of time-to-space ghost imaging. In the process of PDC with pulsed pump in a nonlinear crystal with nonlinearity $\chi$, two entangled photons are generated at phase-matched directions with anti-correlated frequencies. Two filters and two mirrors select part of the PDC radiation, corresponding to a linear angle-frequency dependence for each beam. The reference beam is far-field imaged on a high-resolution camera without temporal resolution, measuring thus the angle of emission of the reference photon. The test beam is collimated into a single-mode fiber and passes through a frequency-to-time mapping system, represented by a dispersive medium, and subsequently through a temporal object, whose transmittivity changes with time. Finally, the test photon is detected by a single-pixel detector, having neither spatial nor temporal resolution. A spatial ghost image of the temporal object is formed by correlating the frames of the camera with the clicks of the single-pixel detector.}
\label{fig:scheme}
\end{figure*}

Strong temporal-spatial correlations between two entangled beams generated in type-I parametric downconversion (PDC) have been known for more than a decade \cite{Gatti09,Horoshko12,Gatti12,Perina15,LaVolpe20,LaVolpe21,Roux21}. These correlations can be harnessed for the new technique which can be called time-to-space ghost imaging.
The principal scheme of the proposed experiment is depicted in Fig.~\ref{fig:scheme}. Its key element is a source of correlated photon pairs based on spectrally and angularly filtered radiation of spontaneous PDC in a second-order nonlinear crystal of length $L$ cut for type-I collinear phase matching. Let us denote by $z$ the direction of pump propagation and by $(x,y)$ the coordinates in the transverse plane, so that the optical axis of the crystal lies in the $yz$ plane. The pump is a Gaussian laser beam with center frequency $\omega_p$ polarized along the $y$ axis.
%
The subharmonic radiation appears as an ordinary wave with the central frequency $\omega_s=\omega_p/2$. Since the mirrors shown in Fig.~\ref{fig:scheme} are implied to be sufficiently narrow in the $y$ direction (orthogonal to the plane of the figure), the whole problem can be considered as two-dimensional \cite{LaVolpe21}. Under this simplification, the positive-frequency part of each wave field (in photon flux units) can be written as \cite{Kolobov99}
\begin{equation}\label{E}
E^{(+)}_\alpha(z,x,t) =\int\int\epsilon_\alpha(z,q,\Omega)e^{ik_{\alpha z}(q,\Omega)z + iqx
-i(\omega_\alpha+\Omega)t} \frac{\mathrm{d}q\mathrm{d}\Omega}{(2\pi)^2},	
\end{equation}
where $q$ is the transverse component of the wave vector in the $x$ direction, $\Omega$ is the detuning from the central frequency, and $\alpha$ takes values $p$ and $s$ for the pump and subharmonic fields respectively. The longitudinal component of the wave vector is defined for the subharmonic as $k_{s z}(q,\Omega)=\sqrt{k_s^2(\Omega)-q^2}$, where $k_s(\Omega)=n_o(\Omega)(\omega_s+\Omega)/c$ is the wave vector modulus, with $n_o(\Omega)$ being the refractive index of the ordinary wave and $c$ the speed of light in vacuum. The expression for $k_{pz}(q,\Omega)$ has a similar form with the only difference that the refractive index seen by the extraordinary wave depends on $q$ \cite{LaVolpe21}. Here and below, the limits of integration can be understood as infinite.

\begin{figure}[ht]
\centering
\fbox{\includegraphics[width=\linewidth]{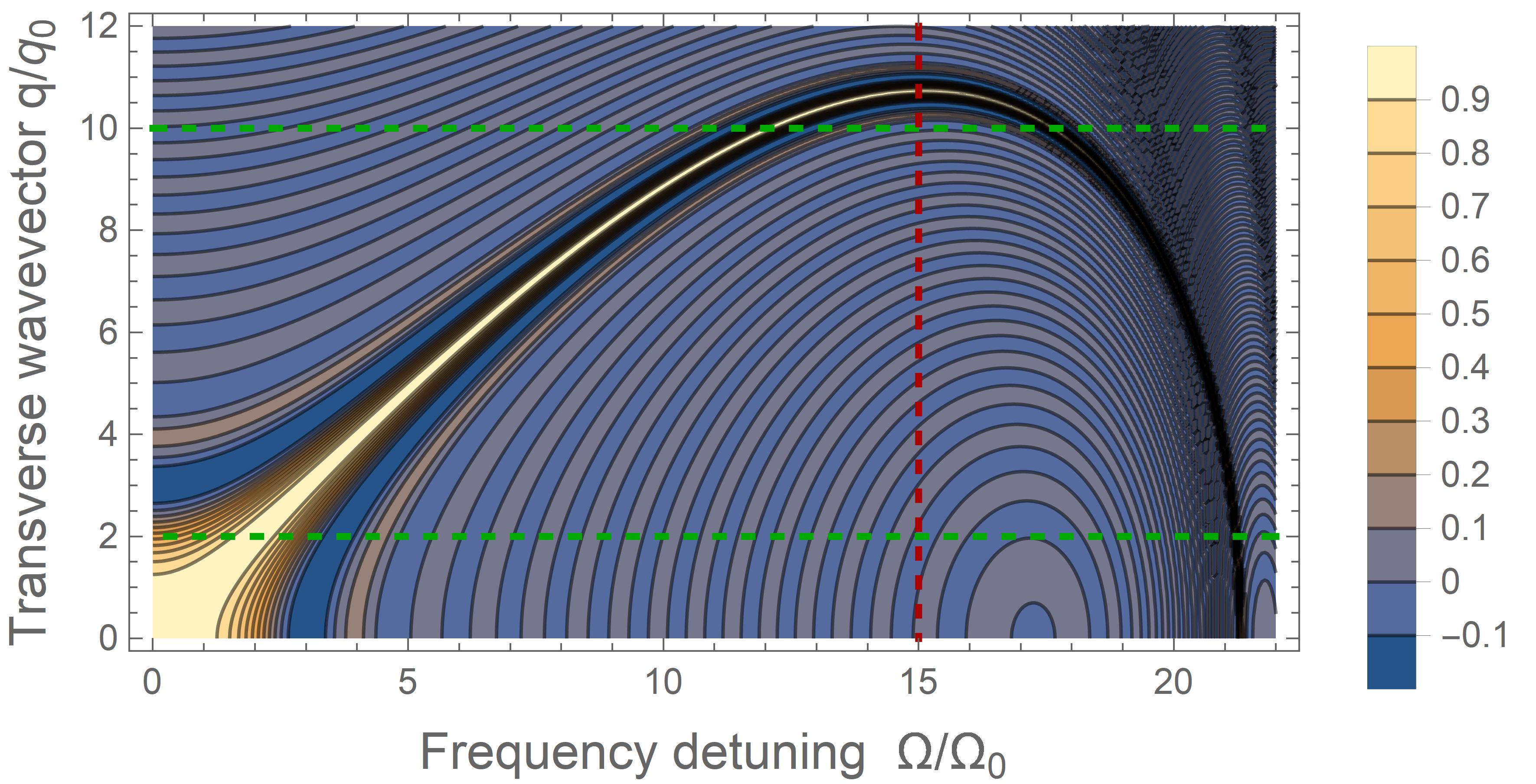}}
\caption{Phase-matching function $\Phi\left(q,\Omega,-q,-\Omega\right)$ of a 5 mm-long BBO crystal pumped at 532 nm. Dashed lines delimit the spectro-angular band selected by the mirrors (green) and the frequency filter (red) for $q_c=6q_0$, $\Delta q=4q_0$, $\Omega_c=15\Omega_0$}.
\label{fig:PMF}
\end{figure}

The pump is assumed to be an undepleted deterministic Fourier-limited Gaussian pulse of full width at half maximum (FWHM) duration $\tau_p$ focused on the input face of the crystal with a waist $w_p$. It is described by a c-function: $\epsilon_p(z,q,\Omega)=\mathcal{E}_p(q,\Omega) =A_0\exp\left(-q^2/4q^2_p-\Omega^2/4\Omega^2_p\right)$,
where $\Omega_p=\sqrt{2\ln2}/\tau_p$, $q_p=1/w_p$, and $A_0$ is the peak amplitude of the pump pulse. However, $\epsilon_s(z,q,\Omega)$ is considered as a photon annihilation operator, satisfying the canonical equal-space commutation relations $[\epsilon_s(z,q,\Omega),\epsilon_s^\dagger(z,q',\Omega')]=(2\pi)^2\delta(q-q')\delta(\Omega-\Omega')$. The evolution of this operator along the crystal is described by the spatial Heisenberg equation \cite{Huttner90}
\begin{equation} \label{Heisenberg}
-i\hbar\frac{\partial\epsilon_s(z,q,\Omega)}{\partial z} = [\epsilon_s(z,q,\Omega),G(z)],
\end{equation}
where the spatial Hamiltonian $G(z)$ is given by the field momentum transferred through the plane $z$ \cite{Horoshko22} and equals
\begin{equation}\label{G}
    G(z) = \chi\int
    \int
    E^{(+)}_p(z,x,t)\left[E^{(-)}_s(z,x,t)\right]^2\mathrm{d}x\mathrm{d}t + \mathrm{H.c.},
\end{equation}
with $\chi$ being the strength of nonlinear coupling and $E^{(-)}_\alpha(z,x,t)=\left[E^{(+)}_\alpha(z,x,t)\right]^\dagger$ being the negative-frequency part of the field. Substituting Eqs. (\ref{E}) and (\ref{G}) into Eq.~(\ref{Heisenberg}) and solving the obtained equation in the first order of perturbation theory for small $\chi$, one obtains the following expression for the subharmonic field at the crystal output
\begin{eqnarray} \label{output}
\epsilon_s\left(L,q,\Omega\right) &=& \epsilon_s\left(0,q,\Omega\right)\\\nonumber
&+&\kappa\int
\int
J\left(q,\Omega,q',\Omega'\right)
\epsilon_s^\dagger\left(0,q',\Omega'\right) \frac{\mathrm{d}q'\mathrm{d}\Omega'}{(2\pi )^2},
\end{eqnarray}
where $\kappa=2i\chi L/\hbar$ is the new coupling constant and
\begin{equation} \label{JSAA}
J\left(q,\Omega,q',\Omega'\right) = \mathcal{E}_p\left(q+q',\Omega+\Omega'\right)  \Phi\left(q,\Omega,q',\Omega'\right)
\end{equation}
is the joint spectro-angular amplitude (JSAA) of two generated photons, which includes the phase-matching function
\begin{equation} \label{Phi}
\Phi\left(q,\Omega,q',\Omega'\right) =e^{-i\Delta\left(q,\Omega,q',\Omega'\right)L/2} \sinc\left(\frac{\Delta\left(q,\Omega,q',\Omega'\right)L}{2}\right),
\end{equation}
depending on the phase mitmatch of the two generated photons
\begin{equation} \label{Delta}
\Delta\left(q,\Omega,q',\Omega'\right) = k_{sz}(q,\Omega)+k_{sz}(q',\Omega')-k_{pz}(q+q',\Omega+\Omega').
\end{equation}
The photon pairs are most efficiently generated at such angles and frequencies that $\left|\Phi\left(q,\Omega,q',\Omega'\right)\right|\approx1$, which implies  $\Delta\left(q,\Omega,q',\Omega'\right)\approx0$. Understanding this process is facilitated in the limit of monochromatic and plane-wave pump, where $\mathcal{E}_p\left(q+q',\Omega+\Omega'\right) \propto\delta\left(q+q'\right)\delta\left(\Omega+\Omega'\right)$ and the phase-matching function can be replaced by $\Phi\left(q,\Omega,-q,-\Omega\right)$. This function is calculated for a beta-barium borate (BBO) crystal from Sellmeier's equations for its refractive indices \cite{Kato86} and shown in Fig.~\ref{fig:PMF}.

The mirrors shown in Fig.~\ref{fig:scheme} provide angular filtering of the subharmonic wave in the far field of the crystal. Therefore, their effect can be taken into account by replacing the JSAA, Eq.~(\ref{JSAA}), by $J_0\left(q_1,\Omega_1,q_2,\Omega_2\right) =J\left(q_1,\Omega_1,q_2,\Omega_2\right)F\left(q_1,\Omega_1\right)F\left(-q_2,-\Omega_2\right)$, where we imply that the arguments $(q_1,\Omega_1)$ refer to the reference beam, the arguments $(q_2,\Omega_2)$ refer to the test beam, and $F\left(q,\Omega\right)$ is the function taking into account the angular and spectral filtering of the reference beam. This function has the form
\begin{equation} \label{filter}
F\left(q,\Omega\right)
=\Pi\left(\frac{\left|q-q_c\right|}{2\Delta q}\right)
\Pi\left(\frac{\left|\Omega-\Omega_c\right|}{2\Omega_c}\right),
\end{equation}
where $\Pi(x)$ is the rectangular function, $\Delta q$ is the half-width of the angular filter, $q_c$ is the central transverse wave vector of the reference beam, while $\Omega_c$ is the central detuning of a frequency filter installed in this arm. The filter function for the test beam is simply $F\left(-q',-\Omega'\right)$, which provides a selection of maximally phase matched directions and frequencies of a photon pair.

The area selected by the filters in the $(q,\Omega)$ space of the reference beam is shown in Fig.~\ref{fig:PMF}. Within this area, we can use the paraxial propagation and quadratic dispersion approximation (PPQDA) \cite{Gatti12,Horoshko12} and write
$k_{sz}\left(q,\Omega\right)\approx k_0+k'_0\Omega+k''_0 \Omega^2/2-q^2/(2k_0)$,
where $k_0=k_s\left(0\right)$, while $k'_0$ and $k''_0$ are the first and second derivatives of $k_s\left(\Omega\right)$ at $\Omega=0$. Substituting this decomposition into Eq.~(\ref{Delta}), we obtain
\begin{equation} \label{Delta0}
\Delta\left(q,\Omega,-q,-\Omega\right)L\approx k^{''}_0{L\Omega}^2-\frac{q^2}{k_0}L=\frac{{\Omega}^2}{{\Omega}^2_0}-\frac{q^2}{q^2_0},
\end{equation}
where $q_0=\sqrt{k_0/L}$, $\Omega_0=\sqrt{1/k''_0L}$ are the characteristic sizes of the central region near the origin of JSAA, where no spectro-angular coupling occurs \cite{Horoshko12,Gatti12}, and a collinear phase-matching, $2k_0=k_{pz}(0,0)$, is assumed. For the model crystal of Fig.~\ref{fig:PMF}, $q_0=44$ rad/mm, $\Omega_0=69$ rad/ps. We see from Eq.~(\ref{Delta0}) that every angle in the reference beam defined by $q>0$ is perfectly phase matched to a frequency of the same beam $\Omega=q\Omega_0/q_0$, which depends linearly on $q$. This linear dependence lies in the heart of the proposed technique.

The ghost image is created by postselecting only those frames of the camera, which are correlated with the clicks of the single-pixel detector \cite{Belinskii94,Pittman95,Bennink02,Gatti04}. For simplicity, we consider only a point-like temporal object, for which the transmittivity equals to 1 in a short temporal window around $t=t_2$ and to 0 at earlier and later time. Thus, the single-pixel detector in the test arm can produce clicks at a time $t=t_2$ only. The frames of the camera postselected on such clicks produce an image determined by the cross-correlation function
\begin{equation} \label{Corr}
C\left(x_1,t_2\right)=\int\int
\left\langle I_R\left(x_1,t_1\right) I_T\left(x_2,t_2\right)\right\rangle
\mathrm{d}t_1\mathrm{d}x_2 ,
\end{equation}
where $I_\beta\left(x,t\right)=E^{(+)}_\beta(z_d,x,t)E^{(-)}_\beta(z_d,x,t)$ is the intensity at the detection plane $z=z_d$ of the reference ($\beta=R$) and test ($\beta=T$) fields. The field at the detection plane is given by a linear integral transformation of the field at the output face of the crystal:
\begin{equation} \label{Ezd}
E^{(+)}_\beta(z_d,x,t) = \int\int h_\beta(x,x')H_\beta(t,t')E^{(+)}_s(L,x',t')\mathrm{d}x' \mathrm{d}t',
\end{equation}
where $h_\beta(x,x')$ and $H_\beta(t,t')$ are the spatial and temporal impulse response functions respectively for the corresponding arm. The filtering is implicit in Eq.~(\ref{Ezd}) by limiting the range of angles and frequencies in each arm to those determined by the filters, as discussed above.

In the reference arm, the camera is placed in the crystal far field in the focal plane of a lens with the focal length $f$, therefore $h_R(x,x')=-ik_0/(2\pi f)\exp\left(-ik_0xx'/f\right)$ \cite{Gatti04} and $H_R(t,t')=\delta(t-t'-t_R)$, where $t_R$ is the delay time in the reference arm. Thus, the field at the detection plane of the reference arm is just a delayed spatial Fourier transform of the subharmonic field in the corresponding spectro-angular band.

In the test arm, all light is collected and directed to the single-pixel detector. The exact form of the impulse response function $h_T(x,x')$ is irrelevant, it is important only that this light does not experience transversal position-dependent delay, so that the transformations in time and space factor out, as implied by Eq.~(\ref{Ezd}). The unitarity property $\int h_T(x,x')h_T^*(x,x'')\mathrm{d}x=\delta(x'-x'')$ is used for integrating Eq.~(\ref{Corr}). The temporal impulse response function $H_T(t,t')$ corresponds to a propagation in a dispersive medium with a group delay $t_T$ and group delay dispersion (GDD) $D_T$. This function is stationary, $H_T(t,t')=H(t-t')$, and, in the quadratic dispersion approximation, its Fourier transform is $\tilde H(\Omega) =\exp\left(iD_T\Omega^2/2+it_T\Omega\right)$ \cite{Patera18}.

The subharmonic field generated in spontaneous PDC obeys Gaussian statistics, which remains Gaussian under linear transformations in both arms. Therefore, the fourth-order correlation function in Eq.~(\ref{Corr}) can be written as a sum of three possible products of second-order correlation functions \cite{Erkmen08}. The phase-sensitive correlation function $\Gamma^{(0,2)}\left(x_1,t_1,x_2,t_2\right)=\left\langle E_R^{(+)}(z_d,x_1,t_1)E_T^{(+)}(z_d,x_2,t_2)\right\rangle e^{i\omega_0(t_1+t_2)}$ has the first order in the smallness parameter $\chi$, while the phase-insensitive correlation functions have second order in this parameter. Therefore, in the low-gain regime of PDC, Eq.~(\ref{Corr}) can be rewritten as
\begin{equation} \label{Corr2}
C\left(x_1,t_2\right) =\int \int
\left|\Gamma^{(0,2)}\left(x_1,t_1,x_2,t_2\right) \right|^2 \mathrm{d}t_1\mathrm{d}x_2.
\end{equation}

\begin{figure}[ht]
\centering
\fbox{\includegraphics[width=\linewidth] {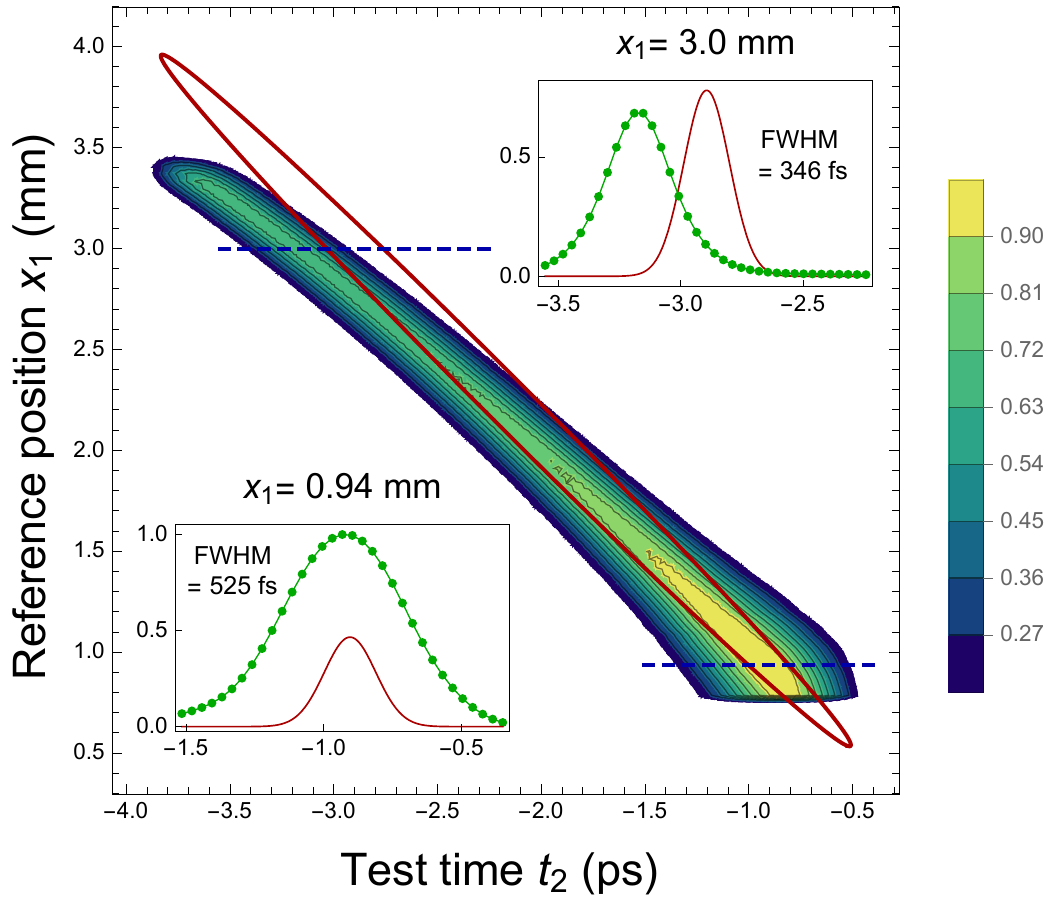}}
\caption{Temporal-spatial cross-correlation function for two beams emitted by the crystal of Fig.~\ref{fig:PMF} pumped by pulses of duration $\tau_p=200$ fs, with $D_T=5233$ fs$^2$ and $f=5$~cm. The color map shows the numerical solution, the solid red line corresponds to the analytical solution being equal to 0.27. The insets show temporal numerical (green dots) and analytical (solid red line) dependencies at positions marked by dashed blue lines.}
\label{fig:C}
\end{figure}

Substituting Eqs.~(\ref{E}), (\ref{output}), (\ref{Ezd}) into Eq.~(\ref{Corr2}) and taking into account the filtering gives
\begin{equation} \label{Corr3}
C\left(x_1,t_2\right) \propto \int Q\left(x_1,\Omega_-\right)e^{-i\Omega_-(t_2-t_T)}
\frac{\mathrm{d}\Omega_-}{2\pi},
\end{equation}
where, in the limit $w_p\to\infty$, implying a plane-wave pump,
\begin{eqnarray} \label{Q}
Q\left(x_1,\Omega_-\right) = \int \int
J_1\left(\frac{k_0x_1}f,\Omega_1,-\frac{k_0x_1}f,\Omega_++\Omega_-/2\right) \\\nonumber
\times J_1^*\left(\frac{k_0x_1}f,\Omega_1,-\frac{k_0x_1}f,\Omega_+-\Omega_-/2\right) e^{iD_T\Omega_+\Omega_-}\frac{\mathrm{d}\Omega_1\mathrm{d}\Omega_+}{(2\pi)^2}
\end{eqnarray}
with $J_1\left(q_1,\Omega_1,q_2,\Omega_2\right)=J_0\left(q_1,\Omega_1,q_2,\Omega_2\right) e^{ik_{sz}(q_1,\Omega_1)L+ik_{sz}(q_2,\Omega_2)L}$. Double integration in Eq.~(\ref{Q}) can be performed numerically with the help of Sellmeier's equations for the dispersion of BBO \cite{Kato86}, and the function $C\left(x_1,t_2\right)$ can be obtained by a fast Fourier transform over the second argument. The results of such a numerical modeling are presented in Fig.~\ref{fig:C}.

An analytical expression for the cross-correlation function, Eq.~(\ref{Corr3}), can be obtained by replacing the JSAA, Eq.~(\ref{JSAA}), with an approximate Gaussian function and taking analytically the integrals in Eqs.~(\ref{Corr3}) and (\ref{Q}). It can be carried out using several approximations \cite{Horoshko19,LaVolpe21}. The first of them is the PPQDA, described above. The second one is the nearly plain-wave and monochromatic pump approximation (NPMPA) based on the assumption that the standard deviations ${\Omega}_p$ and $q_p$ of the pump are so small that the dependencies on $q+q'$ and $\Omega+\Omega'$ can be neglected in the phase-matching and filtering functions. The third one consists in replacing the phase-matched area selected by the filters by a Gaussian function
\begin{equation} \label{Phi2}
\Phi\left(q,\Omega,q',\Omega'\right)\approx \exp\left[-\frac{\nu^2_-}{2\sigma^2_\nu} -\frac{\left(\mu_--\mu_c\right)^2}{2\sigma^2_\mu}-i\mu_c\nu_-\right],
\end{equation}
where $\mu =\left(\Omega/\Omega_0+q/q_0\right)/\sqrt{2}$ and $\nu =\left(\Omega/\Omega_0-q/q_0\right)/\sqrt{2}$ are the coordinates in a rotated basis, $\mu_c=\sqrt{2}q_c/q_0$, and $\mu_-=(\mu-\mu')/2$ with a similar definition for $\nu_-$. The standard deviations of the double Gaussian are $\sigma_\mu=\sqrt{2}\Delta q/q_0$ and $\sigma_\nu=\sigma_s/\sqrt{2}\mu_c$, where $\sigma_s=1.61$ is given by a Gaussian model of $\sinc(x)\approx e^{-x^2/2\sigma^2_s}$ \cite{Horoshko19}. Such a Gaussian modeling gives in the limits $w_p\to\infty$ and $\sigma_\mu\gg\sigma_\nu,\sigma_p$ ($\sigma_p=\Omega_p/\Omega_0$)
\begin{equation} \label{Corr4}
C\left(x_1,t_2\right) \propto \exp\left[-\frac{(D\bar{x}+\tau)^2}{2\Sigma_\tau^2} -\frac{2(\bar{x}-q_c/q_0)^2}{\sigma_\mu^2}\right],
\end{equation}
where $D=D_T\Omega_0^2$, $\bar{x}=k_0x_1/(fq_0)$, $\tau=(t_2-t_T)\Omega_0$, and
\begin{equation} \label{Sigma}
\Sigma_\tau^2
=\frac14\left(\frac1{4\sigma_\nu^2}+\frac1{\sigma_p^2}\right) \left(1+4D^2\sigma_\nu^2\sigma_p^2\right)
\end{equation}
is the dispersion of the dimensionless time $\tau$.

We see in Fig.~\ref{fig:C}, that for a 200 fs pump, where the NPMPA is still valid \cite{Gatti12,Horoshko12},
the agreement between the numerical and analytical solutions is fairly good. The relation between the object time $t_2$ and the measured position $x_1$ is almost linear, except at high $x_1$, which corresponds to high $|\Omega|$, where the PPQDA starts to fail. The precision of time measurement is determined by the width of the cross-correlation function, which  lies within 350-530 fs. We see from Eq.~(\ref{Corr4}) that the dynamical range for $\tau$ grows with the GDD of a dispersive medium used in the test arm. However, as Eq.~(\ref{Sigma}) shows, this GDD also increases the resolution time $2\Sigma_\tau$, when $2D\sigma_\nu\sigma_p$ becomes comparable with 1. In our example $2D\sigma_\nu\sigma_p=0.57$ and $2\sigma_\nu>>\sigma_p$, therefore, $2\Sigma_\tau\approx 1/\sigma_p$ and the temporal resolution is mainly determined by the duration of the pump pulse.

In summary, its has been shown that the time-to-space ghost imaging technique allows one to improve the resolution time of TGI by more than two orders of magnitude, reaching a sub-picosecond-scale resolution. It should be noted, that a recently proposed technique of computational TGI (without reference arm) \cite{Zhao21} reaches even better temporal resolution at the test wavelength 800 nm, but does not admit a dual-color scenario.

This work is funded by Agence Nationale de la Recherche, France, grant ANR-19-QUANT-0001 (QuICHE).

\bibliography{TemporalGI-2023}
\end{document}